\documentclass[twocolumn,showpacs,preprintnumbers]{revtex4}
\usepackage{amssymb}
\usepackage{amsfonts}
\usepackage{amsmath}
\usepackage{graphicx}
\usepackage{dcolumn}
\usepackage{bm}

\input{tcilatex}

\begin{document}

\title{Electron Spin Dynamics of the Superconductor CaC$_{6}$ probed by ESR }
\author{F. Mur\'{a}nyi, G. Urbanik, V. Kataev and B. B\"{u}chner }
\affiliation{Leibniz Institute for Solid State and Materials Research Dresden, 01171
Dresden, PO BOX 270116, Germany}
\date{\today }

\begin{abstract}
Conduction Electron Spin Resonance (CESR) was measured on a thick slab of CaC%
$_{6}$ in the normal and superconducting state. Non-linear microwave
absorption measurements in the superconducting state describe CaC$_{6}$ as
an anisotropic BCS superconductor. CESR data in the normal state
characterize CaC$_{6}$ as a three-dimensional (3D) metal. The analysis
suggests that the scattering of conduction electrons is dominated by
impurities and supports the description of superconductivity in the dirty
limit. A surprising increase of the CESR intensity below $T_{c}$ can not be
explained by the theoretically predicted change in spin susceptibility. It
is interpreted as a vortex enhanced increase of the effective skin depth.
The study of the spin dynamics in the superconducting state and the
discovery of the vortex enhanced increase of the skin depth poses a
challenge to theory to provide a comprehensive description of the observed
phenomena.
\end{abstract}

\pacs{74.70.Ad, 74.25.Nf, 76.30.Pk, 74.25.-q}
\maketitle

\address{
}%

\begin{center}
\textbf{I. INTRODUCTION}
\end{center}

Electron spin dynamics in the superconducting state is a barely studied
field. It was predicted some time ago \cite{Kaplan}\cite{deGennes} that
Conduction Electron Spin Resonance (CESR) can be measured in the
superconducting state. A decrease of the spin susceptibility, which together
with the skin depth determine the CESR intensity, below $T_{c}$ was
predicted by Yosida \cite{Yosida} in the case of no magnetic field. The
temperature dependence of the electron relaxation time, $T_{1}$, was
calculated under the same conditions\ and predicts an increase of $T_{1}$
below $T_{c}$ \cite{Yafet_sc}. However, to date, only a few experimental
works on CESR in superconductors are available. Vier and Schultz \cite%
{VierSchultz} observed CESR on thin Nb foils below $T_{c}$. They found an
increase in $T_{1}$ and a significant decrease in the intensity below $T_{c}$%
. CESR has been reported for K$_{3}$C$_{60}$ powder by Nemes et al. \cite%
{Nemes}. They observed a sudden decrease in intensity below $T_{c}$ and an
increase in $T_{1}$. CESR on the powder of the\ two-gapped superconductor,
MgB$_{2}$, have also been studied \cite{SimonMgB2}. The measured intensity
decrease below $T_{c}$ but the two-gap nature of MgB$_{2}$ makes proper
analysis difficult. CESR also provides direct access to the conduction band
properties of metals and conducting compounds such as intercalated
fullerenes \cite{Nemes}\cite{SimonNH3} and intercalated graphite \cite%
{Lauginie}.

Recently, a graphite intercalated compound (GIC), CaC$_{6}$ has attracted
great attention due to the discovery of superconductivity in YbC$_{6}$ and
CaC$_{6}$ \cite{weller_nature} with transition temperatures, $T_{c}$, of 6.5
K and 11.5 K, respectively. Superconductivity in GICs was first reported in
the potassium-graphite donor compound KC$_{8}$ with $T_{c}$ of 0.14 K \cite%
{Hannay_KC8}. Until the discovery of superconductivity in YbC$_{6}$ and CaC$%
_{6}$ the highest transition temperature was obtained at KTl$_{1.5}$C$_{5}$
with T$_{c}$ of 2.7 K among all the GICs synthetized at ambient pressure 
\cite{wachnik}. However, using high pressure synthesis $T_{c}$ can increase
to 5 K in metastable compounds such as NaC$_{2}$ and KC$_{3}$ \cite{Belash}%
\cite{Avdeev}. Despite its much higher $T_{c}$, CaC$_{6}$ can also be
described as a classical BCS superconductor with an isotropic gap. The zero
temperature gap value is\ $\sim $1.7 meV \cite{STSpaper}\cite{LamuraPRL}.
The upper critical field, $H_{c2}$, shows a remarkable anisotropy with the
zero temperature value changing between $\sim $0.4 T and $\sim $1.9 T,
depending on the external magnetic field direction \cite{critical_field}.

The scarcity of CESR observations in the superconducting state and the
recent discovery of CaC$_{6}$ have motivated the present work. Our aims
were: i) to obtain insight into the superconducting state of this novel
material ii) to study the normal state properties by directly probing the
electron spin system by CESR. A striking result of our study is the
observation of a surprising increase of the CESR intensity below $T_{c}$,
which can not be explained by changes in the spin susceptibility. We
interpret this observation as an enhancement of the microwave skin depth via
vortex dynamics. Our non-linear microwave absorption measuments in the
superconducting state support the description of CaC$_{6}$ as an anisotropic
BCS superconductor. Further, our results on CESR in the normal state
characterize CaC$_{6}$ as a 3D metal and strongly suggest that the
superconductivity can be described in the dirty limit.

\begin{center}
\textbf{II. EXPERIMENT}

\bigskip

A) Sample preparation
\end{center}

CaC$_{6}$ samples were prepared from Highly Oriented Pyrolytic Graphite
(HOPG, Stucture Probe Inc., SPI-2 Grade) and calcium metal (Sigma Aldrich,
99.99\% purity). Because the vapour transport method yields only
superficially intercalated samples \cite{weller_nature} we followed the
preparation method of N. Emery \textit{at al.} \cite{Emery_synthesis}.
Lithium and calcium are strongly water and air sensitive, therefore the
lithium-calcium alloy with atomic ratio of 4:1 was prepared in a high purity
(O$_{2}$ \TEXTsymbol{<} 1 ppm, H$_{2}$O \TEXTsymbol{<} 1 ppm) argon-filled
glove-box. 1 g calcium and 0.7 g lithium (Sigma Aldrich, 99.9\% purity) were
measured in the stainless steel reactor. The reactor was put on a hot plate
in the glove-box. Lithium melts at 180.6 $^{0}$C and wets and dissolves
calcium pieces. The alloy's melting point is approximately 230 $^{0}$C at
this atomic ratio. Removal of remaining dirt from the alloy and the mixing
of two components were carried out by a small stainless steel spoon. Due to
the high surface tension of the prepared alloy the dirt floats on the
surface. HOPG pieces were immersed into the molten alloy with the help of
thin tantalum bands. The upper part of the graphite holder fits into the
groove formed on the top of reactor holding the graphite in the alloy,
preventing the emergence. Another groove was made with a cutting edge inside
for sealing. After cooling down the reactor was tightly closed with a copper
gasket because lithium vaporizes from the molten alloy and the alloy also
captures the remaining oxygen present even in a high purity glove-box. 
\begin{figure}[th]
\includegraphics[width=1\hsize]{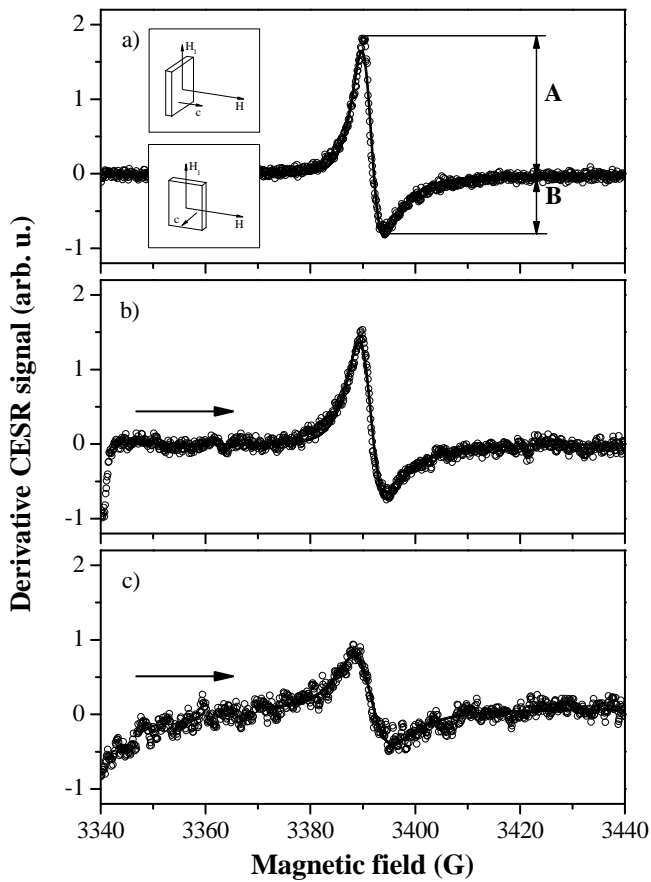}
\caption{Selected CESR spectra of CaC$_{6}$ recorded at 12.2, 6 and 3.75 K, 
\textit{a}, \textit{b}, \textit{c}, respectively, H$\parallel $\textit{ab}%
-plane. Open circles represent measured curves, solid lines are Dysonian
fits. $A/B$ is the asymmetry parameter, ratio of the maximum, $A$, and the
minimum, $B$. The arrows indicate the magnetic field sweep direction. Insets
show the two magnetic field arrangements: H$\Vert $\textit{c}-axis and H$%
\Vert $\textit{ab}-plane. $H_{1}$ is the magnetic component of the
microwave. For detailed explanation see the text.}
\label{diam1}
\end{figure}
The closed reactor was placed and sealed once more in a stainless steel tube
equipped with an other copper gasket. The 10 days long intercalation at 350 $%
^{0}$C was carried out in an industrial furnace to achieve homogeneous
temperature distribution. After the intercalation the reactor was opened and
heated up on a hot plate in the glove-box. Removed samples were cleaned by
sharp knife and sandpaper. The sample's color is gold and the surface
possesses a shiny, metallic lustre. The homogeneity of intercalation was
checked by scratching the sample from one side to the opposite one. Every
slice revealed the same color and brightness, no clue of unintercalated
region was found.\ To check the superconductivity, magnetization
measurements were conducted in a SQUID magnetometer.

\begin{center}
B) ESR measurement
\end{center}

ESR measurements were carried out with an X-band (9.5 GHz) spectrometer
(Bruker EMX). The sample choosen for ESR measurement with dimensions of 2 x
2.3 x 0.3 mm$^{3}$ was glued on to a teflon sample holder with a tiny
droplet of vacuum grease (Apiezon N-type) and sealed in a quartz tube under
argon atmosphere.

CESR is an efficient tool to characterize GICs because of its sensitivity to
the electron spin dynamics and diffusion \cite{Lauginie}\cite{Dresselhaus}.
The analysis of the data was carried out in the framework of the classical
Dyson theory \cite{Dyson}\cite{Feher}. Fig. 1. shows selected CESR spectra,
field derivative of the absorption, $dP(H)/dH$, recorded at applied magnetic
field, $H$, in the \textit{ab}-plane. The inset shows the two magnetic field
arrangements: H$\Vert $\textit{c}-axis and H$\Vert $\textit{ab}-plane. The
observed lineshape is Dysonian over the whole temperature range studied and
is typical of metallic conductors. The resonance field, $H_{0}$, the
linewidth, $w$, and the intensity , $I$, are resulted from Dysonian fits
(solid curves). The ratio of the two extrema, $A/B$, is the so called
asymmetry parameter \cite{Dyson}, it depends on $\sqrt{T_{D}/T_{2}}$, where $%
T_{D}$ is the diffusion time, that takes an electron to diffuse through the
skin depth, and $T_{2}$ is the spin coherence time. For the detailed
dependence see Fig. 7. in Ref. \cite{Feher}. Generally the homogeneous CESR
linewidth measures the spin coherence time, $T_{2}=1/\gamma _{e}w$, where $%
\gamma _{e}$\ is the electron gyromagnetic ratio. However, in light metals, $%
T_{2}$, and the spin lifetime, $T_{1}$, are equal \cite{Yafet_T1} and
therefore the linewidth measures the spin lifetime, $T_{1}$, too. Panel 
\textit{b)} and \textit{c)} of Fig. 1. show the evolution of the CESR line
in the superconducting state. The occurance of supercunductivity is
indicated by the non-linear start of the magnetic field sweeps and by the
enhancement of the noise level due to the so-called vortex noise generated
by the magnetic field modulation in the superconducting state below the
irreversibility line \cite{Janossyvortex}.

\begin{center}
\textbf{III. RESULTS AND DISCUSSION}

\bigskip

A) Nonlinear absorption
\end{center}

In the superconducting state the magnetic field dependent nonlinear
absorption \cite{nonlinmwabs}\cite{nonlinmwabs2} was observed (Fig. 2.). The
microwave losses in the superconducting state in the presence of the
magnetic field arise due to Lorentz force driven fluxon dynamics. 
\begin{figure}[th]
\includegraphics[width=1\hsize]{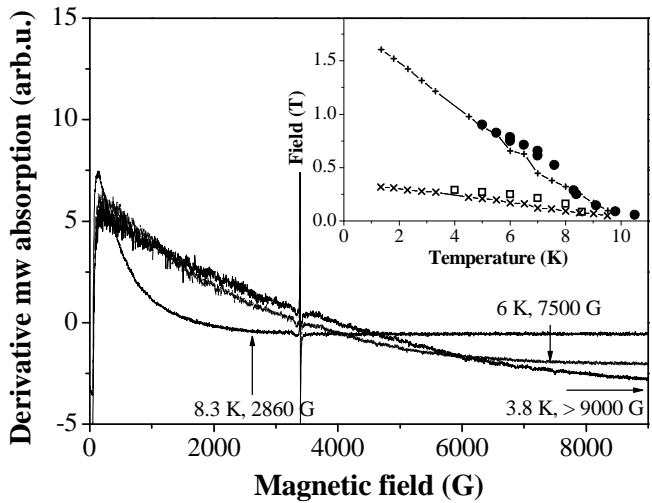}
\caption{Non-linear microwave absorption in the superconducting state at
3.8, 6 and 8.3 K, H$\parallel $\textit{ab}-plane. Arrows indicate the upper
critical field, $H_{c2}$. Note the sharp signal at 3390 G which is the CESR
of CaC$_{6}$. Inset shows the measured $H_{c2}$ values. Open squares and
full circles denote H$\parallel $\textit{c}-axis and H$\parallel $\textit{ab}%
-plane, respectively. $\times $ and $+$ are measured $H_{c2}$ values in Ref.%
\protect\cite{critical_field}.}
\end{figure}
The non-linearity in the microwave absorption disappears when the external
magnetic field suppresses the superconductivity. This provides the
possibility to estimate the upper critical field, $H_{c2}$. The results for
both magnetic field orientations are shown in Fig. 2. inset. The data reveal
a linear dependence of $H_{c2}(T)$ with an appreciable anisotropy and thus
support the magnetoresistance results of Ref.\cite{critical_field}. Both
data sets agree quantitatively well (Fig. 2. inset) and support the
description of CaC$_{6}$ in terms of an anisotropic BCS superconductor.

\begin{center}
B) CESR in the normal state
\end{center}

The CESR results are shown on Fig. 3. The temperature dependence of the $g$%
-factor, the linewidth, the asymmetry parameter and the signal intensity
were studied in the temperature range 3.8 K $\leq T\leq $ 300 K. In the
normal state the $g$-factor of CaC$_{6}$, Fig. 3.a, was measured with an
Mn:MgO reference sample with the $g$-factor of 2.0009 ($\hbar \omega =g\mu
_{B}B_{0}$, where $\hbar $ is the reduced Planck constant, $\omega $\ is the
angular frequency, $\mu _{B}$\ is the Bohr magneton and $B_{0}$ is the
resonance field). The $g$-factor of the mother compound, graphite, is highly
anisotropic with $g_{c}$ = 2.0496 and $g_{ab}$ = 2.0026 \cite{Wagoner}. The
expectation that intercalation may possibly increase the 2-dimesionality of
the mother compound due to the increase of the interlayer distance is not
supported by the $g$-factor values. The measured $g$-factor, $g_{c}$ = $%
g_{ab}$ = 1.9987(5), is orientation and temperature independent below 220 K,
which is a characteristic of metals and usual among GICs \cite{Lauginie}.
The isotropic and temperature independent $g$-factor supports the assumption
that Ca intercalation does not play the role of a simple spacer but creates
bonds between the graphene sheets thus increasing the 3D-character of the
band structure. A slight anisotropy of the $g$-factor can be observed above
220 K.

The linewidth in both orientatons is temperature independent and slightly
orientation dependent (Fig. 3.b). The calculated spin lifetime for $w\sim $
4 G is $T_{1}=T_{2}=1/\gamma _{e}w=$ 14.2 ns. The temperature independent
behavior of $T_{1}$\ means that the spin relaxation is mostly governed by
impurities and scattering processes are dominated by disorder. No clear
signature of temperature dependence i.e. phonon contribution is present.
This justifies the assumption that the superconductivity in CaC$_{6}$ can be
described in the dirty limit \cite{LamuraPRL} and agrees with the general
specifics of the\ intercalation process, which includes the presence of a
certain amount of disorder in the light metal (in this case: Ca)
distribution \cite{Dresselhaus}. The linewidth shows a slight anisotropy
which decreases with decreasing temperature. This observation and the small
anisotropy of the $g$-factor above 220 K could be interpreted as consequence
of a weak anisotropic thermal contraction. It is worth to mention that the
Elliott-Yafet relation, $\tau /T_{1}\sim \Delta g^{2}$, \cite{Yafet_T1}\cite%
{Elliott}\cite{Monod} which connects the momentum lifetime, $\tau $, the
spin lifetime, $T_{1}$, and the deviation from the free electron $g$-factor, 
$\Delta g$, in light metals does not hold in CaC$_{6}$.

The asymmetry parameter, $A/B$ (Fig. 3.c), is related to the ratio of $T_{2}$
and the diffusion time, $T_{D}$, of an electron through the skin depth \cite%
{Dyson}\cite{Feher}. In particular, for asymmetries larger than $\sim $3.5, $%
T_{D}$ gets comparable and even shorter than $T_{2}$. However, in CaC$_{6}$
the $A/B$ values lie in the range 1.7-3.3 implying that $T_{D}$ is much
longer than the spin coherence time, $T_{2}$. The electron relaxes and
looses its spin memory in the skin depth. This conclusion agrees with the
presence of disorder in this material (dirty limit). We note that the $A/B$
parameter goes below the theoretical limit in the \textit{c}-direction. Also
in contrast to the theoretical expectation, the A/B ratio slightly decreases
with decreasing temperature in spite of the slight increase of the
penetration depth. These discrepancies might arise from the layered
structure of the\ studied material, whereas Dyson's theory was developed for
an isotropic metal. The layered structure of CaC$_{6}$ slightly reflects in
the anisotropy of the\ asymmetry parameter but its values lie in such a
range that drawing additional conclusions regarding the diffusion time are
not possible.

The temperature dependence of CESR signal intensity, $I(T)$, is determined
by the spin susceptibility of conduction electrons and a microwave skin
depth, $\delta =\sqrt{\rho /\pi f\mu _{0}}$ where $\rho $ is the
resistivity, $f$ is the microwave frequency (9.5 GHz in our case) and $\mu
_{0}$ is the vacuum magnetic permeability. 
\begin{figure}[th]
\includegraphics[width=1\hsize]{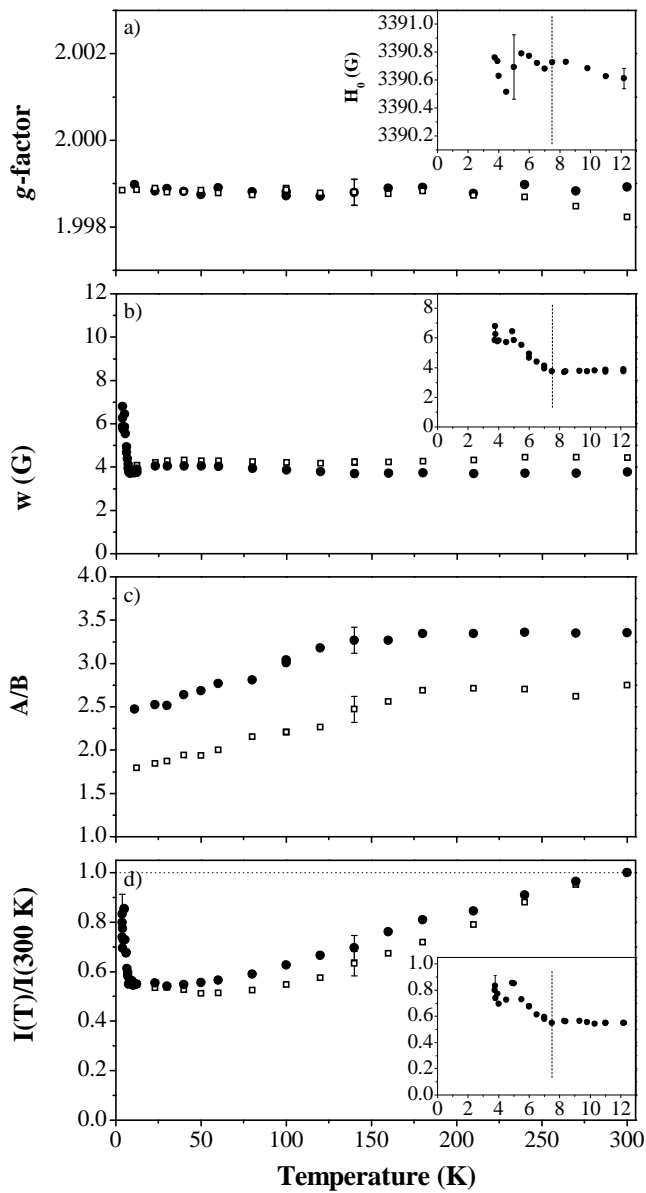}
\caption{The temperature dependence of a) $g$-factor, b) linewidth, $w$, c)
asymmetry parameter, $A/B$ and d) signal intensity, $I$. Open squares and
full circles denote H$\parallel $\textit{c}-axis and H$\parallel $\textit{ab}%
-plane, respectively. The insets show the resonance field, the linewidth and
the intesity in the superconducting state. Dotted lines indicate the
superconducting transition.}
\label{diam3}
\end{figure}
Fig. 3.d shows the temperature dependence of the normalized intensity, $%
I(T)/I(300K)$. The intensity was measured with a ruby reference sample
(National Bureau of Standards, SRM-2601) placed outside the cryostat and
glued on to the cavity wall to keep it at a constant temperature. With the
resistivity data of Ref.\cite{critical_field} and \cite{Gauzzi} the
calculation yields a skin depth at 300 K in the \textit{c}-direction $\delta
_{c}$ = 5.5 $\mu m$\ ($\rho =$ 115 $\mu \Omega cm$ ) and in the \textit{ab}%
-plane $\delta _{ab}$ = 3.5 $\mu m$ ($\rho =$ 46 $\mu \Omega cm$\ ). From
300 K down to $T_{c}$ the intensity decreases by a factor of 2 and there is
no significant difference between the two magnetic field orientations. The
sample geometry yields that $\delta _{c}$ is relevant for the determination
of the volume fraction accessible to microwaves (Fig. 1. inset). Taking into
account the temperature dependence of the resistivity in the \textit{c}%
-direction \cite{critical_field} it can be concluded that the measured
intensity changes in accordance with the change of the skin depth implying
that the spin susceptibility is Pauli-like temperature independent. The air
sensitive nature of CaC$_{6}$ did not allow its proper mixing with an
intesity standard for the measurement of the absolute intensity, so we can
only give an order-of-magnitude estimation. Taking into account the skin
depth the spin susceptibility is $\chi _{S}\sim 10^{-4}$ emu/mol and
assuming negligible electron-electron correlations the density of states at
the Fermi-level is $\sim $3 states/eV. Comparing with theoretical
calculation, $\sim $1.5 states/eV \cite{Kim}, this estimation is reasonable.

\begin{center}
C) CESR in the superconducting state
\end{center}

In the superconducting state both $w$ and $I$ show a strong increase below $%
T_{c}$ whose origin is obviously related to superconductivity. The critical
temperature, $T_{c}$, is shifted to 7.5 K at 3390 G (the resonance magnetic
field at 9.5 GHz) with H$\Vert $ab-plane (Fig. 2. inset). In the other
direction, H$\Vert $c-axis, these phenomena are not observed since the
applied magnetic field suppresses the supercunductivity. The resonance
field, $H_{0}$ (Fig. 3. a inset), does not change within experimental
percision by entering the superconducting state. Existing theories \cite%
{Yosida}\cite{Yafet_sc} do not predict a strong increase of $w$ and $I$. In
the superconducting state of CaC$_{6}$ the onset of the\ irreversibility
close to $T_{c}$, which is a characteristic of a dirty superconductor,
results in an appreciable inhomogeneous broadening of the CESR signal (Fig.
3. b inset). Below the irreversibility temperature the vortex distribution
depends on the thermal and magnetic field history \cite{VierSchultz}\cite%
{Nemes} and causes an inhomogeneous magnetic field inside the sample which
makes the determination of $T_{1}$ from the linewidth\ impossible. The
magnetic field inhomogeneity arising from the presence of vortices is
averaged out because within the lifetime the conduction electrons travel
long distances compared to the distance between vortices. The resonance
field in the superconducting state in case of MgB$_{2}$ is shifted to higher
magnetic fields due to diamagnetic shielding currents \cite{SimonMgB2}. The
diamagnetic shift is decreasing in increasing magnetic field (see Fig. 2. in
Ref. \cite{SimonMgB2}), at $H_{0}$ = 8.1 T ($\sim $225 GHz) it is only $\sim 
$3 G \cite{unpub}.\ Comparing the ratio of $H_{0}$ and the upper critical
field of MgB$_{2}$ in the ab-plane, $H_{0}/H_{c2}^{ab}\sim $0.5 ($%
H_{c2}^{ab}\sim $16 T \cite{mgb2esr}) with that in the present work $%
H_{0}/H_{c2}^{ab}\sim $0.3 and taking into account that the diamagnetic
shift scales with the upper critical field its absence in CaC$_{6}$ is
reasonable. Surprisingly, the signal intensity (Fig. 3. d inset) starts to
increase below $T_{c}$ in contrast with the theoretical prediction that the
spin susceptibility decreases below $T_{c}$ \cite{Yosida}. A similar
behavior has been reported from\ MgB$_{2}$ samples in Ref.\cite{Koseoglu}.
However, the authors related the intensity increase below $T_{c}$ to
paramagnetic centers which show Curie-like spin susceptibility in the normal
state. This can not be the explanation in our case because CaC$_{6}$ shows
temperature independent (Pauli) spin susceptibility in the normal state \cite%
{paramimp}. A possible explanation is based on the temperature and magnetic
field dependence of the skin depth because the CESR intensity on a thick
sample is also proportional to the skin depth. An increase of the skin depth
as a function of magnetic field has previously been reported on YBa$_{2}$Cu$%
_{3}$O$_{y}$ crystals in the MHz frequency regime \cite{YBCO}. Qualitatively
we can relate the observed increase of the CESR\ intensity to the increase
of the effective skin depth in the superconducting state due to the presence
of vortices \cite{Brandt}\cite{CoffeyClem}. The applied microwave field
interacts with the vortices at the surface and exerts forces on the vortex
ends. The evolving deformation of the vortex lattice then propagates into
the interior, driven forward by the elasticity of the vortex lattice and is
slowed down by pinning and viscous resistance. The build up of the vortex
lattice with decreasing temperature yields an increasing effective skin
depth. However, a proper description of the observed enhancement of the skin
depth calls for improved theory to account for\ the magnetic field
arrangement used in ESR spectrometry (large static magnetic field,
perpendicular mw perturbation, paralell rf magnetic field modulation).

\begin{center}
\textbf{IV. CONLCLUSION}
\end{center}

In conlcusion, we observed and studied Conduction Electron Spin Resonance in
the normal and superconducting state of CaC$_{6}$. The intercalation of
graphite with calcium yields a more 2D crystal structure. However,
electronically CESR characterizes CaC$_{6}$ as a 3D metal thus suggesting
that Ca provides appreciable bonding between the graphene sheets along the 
\textit{c}-axis. The temperature independent behavior of the spin lifetime, $%
T_{1}$, and the asymmetry parameter, $A/B$, provide evidence that the
scattering processes of conduction electrons are dominated by impurity
scattering and supports the description of superconductivity in the dirty
limit. Non-linear microwave absorption measurements in the superconducting
state supports the description of CaC$_{6}$ as an anisotropic BCS
superconductor. CESR data in the superconducting state indicate the
vortex-enhancement of the effective skin depth below $T_{c}$, though the
quantitative data analysis in the superconducting state requires the
extension of theory to the magnetic field arrangement used in ESR
spectroscopy. The study of the spin dynamics in the superconducting state of
CaC$_{6}$ and the observation of the vortex enhanced increase of the skin
depth calls for theoretical works to provide a comprehensive description of
the observed phenomena.

\begin{center}
\textbf{ACKNOWLEDGEMENT}
\end{center}

Part of this work was supported by German Research Foundation (DFG 436 UNG
17/4/05). F. M. greatfully acknowledges the fellowship of the Alexander von
Humboldt Foundation.


\end{document}